# Bending effects and optical properties of $WSe_2$ nanoribbons of topological phase

Hong Tang[*], Jason M. Breslin, Li Yin, and Adrienn Ruzsinszky[†]

Department of Physics, Temple University, Philadelphia, PA 19122

**ABSTRACT** A $WSe_2$ monolayer of 1T′ phase is a large band gap quantum spin Hall insulator, supporting dissipationless charge and spin transports through the topologically protected edge states. In this work, we explore the nanoribbon forms of 1T′ phase $WSe_2$ by first-principles density functional calculations and the many-body perturbation GW and Bethe-Salpeter equation method. We found that the 1T′ $WSe_2$ nanoribbon can show topological edge states with a ribbon width of ~4-6 nm. Those edge bands show crossing through the Fermi level an odd number of times, with one kind of spin-polarization connecting the valence band continuum and conduction band continuum. The topological features of the edge bands hold even under small and medium bending in the nanoribbon, while large bending induces large band splitting, resulting in a topological switch-off in the edge bands. The semiconducting 1T′ $WSe_2$ nanoribbon shows a large tunability with bending in optical absorption spectra and exciton states. The lowest-energy exciton is changed from optically dark in the flat nanoribbon to bright in the bent nanoribbons. These properties in the 1T′ $WSe_2$ nanoribbons suggest potential applications in controllable quantum electronics and exciton-based quantum information processes.

Since the theoretical prediction of the quantum spin Hall (QSH) effect in the two-dimensional (2D) transition metal dichalcogenide (TMD) materials in the 1T′ phase [1], there have been great interests and efforts in experimental and theoretical research on the topological properties of the related TMD and other 2D layered materials [2-19]. Many of the topological TMD monolayers have been synthesized. For examples, Naylor et al. [2] successfully synthesized the 1T′ $MoTe_2$ single layer crystal by the chemical vapor deposition on $SiO_2$/Si substrates and observed the weak antilocalization effect in it, suggesting the confirmation of its topological property. Tang et al. [3] synthesized the topological 1T′ $WTe_2$ monolayer on graphene substrates by molecular beam epitaxy and verified its band inversion, bulk gap (~55 meV) and conducting edge states with angle resolved photoemission and scanning tunneling spectroscopy. By using the same experimental techniques, Chen et al. [4] synthesized the topological 1T′ $WSe_2$ monolayer and determined its large bulk QSH gap of 129 meV. Also, Ugeda et al. [5] synthesized a $WSe_2$ single layer coexisting 1T′ and 1H phases on the SiC substrate and observed the topological states at the crystalline phase boundary. The relatively large bulk band gap (~100 or > 100 meV), scalable edge conducting channels by multilayer stacking, and trivial/nontrivial phase on-off switching by externally applied controls, such as an electric field [1, 6], electrostatic doping [7], or terahertz-field induction [8, 9], make TMD layered materials an attractive platform for quantum electronic devices and related study of new physics.

In a TMD monolayer of 1T′ phase, in which the metal and nonmetal atoms form distorted octahedra, the transition metal atoms, such as W or Mo atoms, form zigzag chains along the short lattice vector of the rectangle 2D unit cell, and cause a band inversion between the metal d-orbital derived band and the chalcogen atom p-orbital derived band around the Γ point in the Brillouin zone (BZ). The strong spin-orbit coupling (SOC) opens a gap between the two inverted bands and forms bulk insulating states in the material, while creating helical conducting edge states in the gap that are topologically protected from backscattering and support dissipationless transports of charge and spin. Dependent on synthesis conditions, the prepared 1T′ phase TMD monolayer, such as $WTe_2$, can show a semimetallic bulk feature, limiting its coherent edge channel length. Strain engineering [10] was shown to be an effective means to tune the bulk electronic structure and transfer the bulk from semimetal to insulating states and hence improve the quantum effect. Strains can also effectively tune the Dirac point [11] and Dirac surface states [12], leading to controllable topological properties.

---

[*] Email: hongtang@temple.edu
[†] Email: aruzsinszky@temple.edu

Nanoribbon is a quasi-one-dimensional form of material and shows many interesting properties. Cao et al. [16] theoretically discovered that graphene nanoribbons (GNR) can support topological phases dictated by unit cell terminations and protected by spatial symmetries. The GNR's topology can be modified by atomic dopants and/or a transverse electric field [17]. The stable spin centers and chains formed by topological junction states in GNR superlattices of distinct topological segments, the existing of topological end states and the controllable coupling between adjacent junction states are among the interesting properties of GNRs [16-19]. The nanoribbons formed from topological 1T′ phase TMD monolayers also have topological edge bands, and the coupling of those edge bands depends on the width of the nanoribbons. Previous study on the TMD nanoribbons revealed the interesting edge magnetic property [20] and rich exciton states [21]. Moreover, bending the nanoribbons can add dramatic modifications to the internal structures and local strains, and hence control their electronic structures and achieve tunable charge localizations and conductivity [22], optical absorptions [21], spin-polarizations [23], and exciton states [21, 23]. In this work, we study the topological properties of 1T′ $WSe_2$ nanoribbons under bending from first-principles density functional theory (DFT) calculations and the optical properties and exciton states from the many-body perturbation GW [24, 25] and Bethe-Salpeter equation (GW+BSE) [26] method. We found that the 1T′ $WSe_2$ nanoribbons can show topological edge states at widths of ~4-6 nm. Those edge bands cross the Fermi level an odd number of times, and each edge band shows one kind of spin-polarization. Small and medium bending in the nanoribbon do not affect the topological features of the edge bands, while large bending usually induces large band splitting and moves more bulk bands to the Fermi level, resulting in a switch-off of the topological features in the edge bands. The optical absorption and exciton states show a large tunability with bending in the semiconducting 1T′ $WSe_2$ nanoribbons. Bending can turn the low-energy dark excitons into optically bright ones. These interesting properties in the 1T′ $WSe_2$ nanoribbons indicate potential applications in controllable quantum electronic nanodevices and exciton-based quantum information processes.

**Results and discussions**

**Topological edge bands and bending effects.** The $WSe_2$ nanoribbons are formed by cutting the pre-relaxed $WSe_2$ monolayer of 1T′ phase, in which the W atoms form zigzag chains along one direction. We form the nanoribbons with the ribbon length direction either parallel or perpendicular to the zigzag W atom chains and denote the former ones as Z$n$ and the later ones as NZ$n$, where Z means zigzag, NZ non-zigzag, and $n$ the number of W atoms in the periodical supercell, i.e., Z8 represents the nanoribbon with its length parallel to the W zigzag chains and having eight W atoms in its supercell. The edges of nanoribbons are passivated with hydrogen atoms to eliminate the dangling bonds. The relaxed structures of nanoribbons under study are shown in Figure 1 (and Figures S1 and S2 in the Supplemental Information (SI)). The supercell vectors a, b and c are aligned with the coordinate axes x, y and z, respectively. The side and top views of the flat Z8 (Z20) $WSe_2$ nanoribbon are shown in Figure 1(a) and (b) (Figures 1(c) and (d)). The side view of Z14 is shown in Figure 1 (e), and its top view of the extended structure in the periodical length direction is shown in Figure 1 (f). In all cases, the ribbon width direction is along the supercell vector a and its length is along the vector c.

The band structures with edge and bulk atom resolutions and spin-polarization resolutions for the flat Z8, Z14 and Z20 are shown in Figure 2. For the relatively narrow Z8 nanoribbon (with a width of 2.3 nm), the two edge W atoms (W1 and W8) mainly contribute to the edge bands around the Fermi level. Among those edge bands, the ones near the Fermi level are labelled as C1, C2, V1, and V2 in Figures 2(a) and (d). The bulk W atoms' contributions are distributed deeply in the valence band continuum (VBC) and conduction band continuum (CBC) and also significantly extend to the bands around the Fermi level (see Figure 2(a)). Bands C1 and C2 are almost energy-degenerate, as are bands V1 and V2. Since the highest occupied band is V1 and it is even-number indexed, bands V2 and C1 are odd-number indexed and C2 is even-number indexed (see Figures 2(e)-(h)). Since the interaction between the two edges is relatively strong in this narrow nanoribbon, there is a band split of about 0.1 eV at the $\Gamma$ point between the edge bands C1 and V1 near the Fermi level. Because of this band splitting, the edge bands cross through the Fermi level even times, and hence are rendered as topologically trivial, i.e., there are no edge bands directly connecting the CBC and VBC. This can be seen clearly from the spin polarization resolved band plots in Figures 2(e)-(h). The spin-polarization produced by the edge atom W1 (see Figure 1 for the locations of the edge atoms), is mainly on the odd-number indexed bands around the Fermi level (Figure 2(e)), while the spin-polarization produced by the other edge atom W8 is mainly on the even-numbered indexed bands (Figure 2(h)). For both edge atom cases (Figures 2(e) and (h)), the spin-up channel (red dots) does not form a continuous conduit connecting CBC and VBC. Instead, it is broken by the gap between C1 and V1, so does the spin-down channel (blue dots).

The situation for the flat Z14 nanoribbon (with a width of 4.1 nm) is slightly different. The edge interaction induced band splitting is significantly reduced because of the wider width. In Figure 2(i), the band splitting between C1 and V1 is about 20 meV around the indicated green points A, B and C. Although the bulk atoms still contribute to the bands around the Fermi level, their contributions are relatively reduced in magnitude (Figure 2(b)). The spin-polarization resolved band structures (Figures 2(j)-(m)) show that the spin-polarization produced by the edge atom W1 remains mainly on the even-number indexed bands V1 and C2, while that of the other edge atom W14 is on the odd-number indexed bands V2 and C1. In Figure 2 (k) for the edge with atom W1, if we apply an electric field along the $(+Z) - \Gamma - (-Z)$ direction, then an electron with a down-spin will evolve from the left to the right in the BZ, starting from V1 in VBC to pass point B, enter C2, pass point A, enter V1, pass point C, enter C2, and end in CBC eventually, keeping the down-spin state in the journey. The journey of the electron crosses through the Fermi level three times and actually connects VBC and CBC, forming a conducting conduit for carriers. Although there are about 20 meV gaps at points A, B and C, the above-mentioned journey of the down-spin electron may have a non-zero probability if the energy cost for flipping the spin of the electron is not small or comparable with the gap of ~20 meV at point A. Similarly, for the same edge with atom W1, the up-spin electron (red spots in Figure 2(k)) has a similar journey when evolving from the right to the left in the BZ from VBC to CBC, compared with the down-spin electron. The situation is similar for the other edge with atom W14 (Figure 2(l)). However, the down-spin (up-spin) electron evolves oppositely in direction in BZ, compared with the down-spin (up-spin) electron in the edge with atom W1. All the journeys of electron mentioned above have one spin polarization and go through from VBC to CBC, showing a spin-momentum locking feature. All these features show that the flat Z14 may be weakly topologically nontrivial, and it is in a weak quantum spin Hall state.

The above observed topologically nontrivial features in Z14 are further pronounced in the flat Z20 (with a width of 5.9 nm), as shown in Figures 2(n)-(r). The band splitting at point A becomes negligibly small ~4meV. The spin polarized edge bands at each edge cross through the Fermi level three times (Figures 2(o) and (r)), each edge band shows one spin-polarization and a spin-momentum locking feature, and bulk atoms contribute even less to the edge bands, showing that Z20 is a quantum spin Hall insulator. The flat Z8, Z14 and Z20 nanoribbons all have time-reversal symmetry, inversion symmetry and mirror symmetry about a plane perpendicular to the ribbon length direction. The significant band splitting at Γ for the narrow Z8 nanoribbon is not due to the SOC effect, and it is related to the quantum confinement induced overlap effect of the wavefunctions of the two edges.

We also assess the changing trend of band structures under different bending for the Z14 nanoribbon. The relaxed bent structures under three different bending curvatures are shown in Figures 1 (g) with the calculated band structures in Figure 3. Figure 3(a) and Figures 3(e)-(i) for the flat Z14 are the same as Figure 2(b) and Figures 2(i)-(m). They are listed for comparison with other bending cases. For the bending curvature R = 36 Å, the atomic layers in Z14 slightly bend upwards toward vector b. The relative positions and bond lengths between atoms are slightly changed (Figure 1(g)). Figure 3(b) shows that the edge and bulk atoms' contributions to the bands around the Fermi level are almost the same as those of the flat case. However, the spin-polarizations produced by the two edges show slightly different features in the bent case. According to Figures 3(k)-(l), the spin-polarizations produced by the edge atom W1 are on both the odd-number and even-number indexed bands around the Fermi level. Especially, within the regions of the two segments BA and AC, the spin-polarizations are mainly on V1 and V2, and only slightly on C1 and C2. The spin-polarizations produced by the edge atom W14 are also both on the odd-number and even-number indexed bands around the Fermi level, see Figures 3(m) and (n). However, within the regions of the two segments BA and AC, the spin-polarizations are mainly on C1 and C2, and to a lesser extent on V1 and V2. Although there is a gap of about 20 meV between C1 and V2 (or V1) at Γ (or point A), a down-spin electron in the edge with atom W1 may still have a non-zero probability to evolve from the left to the right in BZ, starting from V2 in VBC to pass point B, enter C1, pass point A, enter V2, pass point C, enter C2 and end in CBC, keeping the down-spin state and crossing the Fermi level three time in the journey, see Figures 3(k) and (l). A up-spin electron will evolve in the opposite direction in this same edge channel. Similarly, for the edge channel with edge atom W14, (Figures 3(m) and (n)), an up-spin electron can evolve from the left to the right in BZ, starting from V1 in VBC to pass point B, enter C2, pass point A, enter C1 and V2, pass point C, enter C1 and end in CBC, also keeping the up-spin state and crossing the Fermi level three time in the journey. A down-spin electron will evolve in the opposite direction in this same edge channel. We note that the spin-momentum locking channel in this edge with atom W14 may be enhanced, since when crossing point A, there is almost no gap between C1 and C2. All those features demonstrate that the Z14 nanoribbon under bending R = 36 Å is still topologically non-trivial.

Under bending R = 12 Å, the nanoribbon is obviously bent towards the positive direction of vector b (Figure 1(g). The lower Se atom layer is compressed, while the upper Se atom layer is stretched. This induces, to some extent, the deformation of the distorted octahedra in the nanoribbon structure, hence the changes in bond lengths and angles. It also induces the increase in SOC energy on the W atoms, as can be seen in Figure S4 in the SI, especially for W atoms with indexes 4, 6, 8, and 10. This results in a further band splitting of the edge bands around the region of segments BA and AC, especially for bands C1 and C2, as shown in Figure 3(o). The edge and bulk atoms' contributions to the bands around the Fermi level are slightly changed, compared to the flat case, with a slight increase of the contribution from the bulk atoms to the bands around the Fermi level, as shown in Figure 3(c). The spin-polarizations produced by the edge atoms W1 and W14 show an asymmetric distribution feature among the bands within the two segments BA and AC. Figures 3(p) and (q) suggest that the edge atom W1 induced spin-polarizations are on bands

V1 and V2 within the two segments BA and AC, while the edge atom W14 induced spin-polarizations remain on bands C1 and C2 within the same two segments, as shown in Figures 3(r) and (s). This different distribution of spin-polarizations among the bands V1, V2, C1 and C2 within the region of the two segments for the bent and flat case compared, is due to the loss of the inversion symmetry of the structure in the bent nanoribbon. For the edge channel with atom W1, see Figures 3(p) and (q), the up-spin electron can evolve from the left to the right in BZ, starting from V1 in VBC, to pass point B, keep in V1, pass point A, enter V2, pass point C, enter C2, and end in CBC, crossing the Fermi level three times, and similarly, the down-spin electron can evolve in the opposite direction in this channel. Similarly, for the edge channel with atom W14 (Figure 3(r) and (s)), the up-spin electron can evolve from the left to the right in BZ, starting from V1 in VBC, to pass point B, enter C2, pass point A, enter C1, pass point C, keep in C1, and end in CBC, crossing the Fermi level three times, and also the down-spin electron can evolve in the opposite direction in this channel. The above-mentioned features show that the Z14 nanoribbon under bending R = 12 Å is also topologically non-trivial.

At R = 9 Å, the large bending dramatically changes the structure of the nanoribbon. Per Figure 1(g), although the two sides of the bent nanoribbon show relatively flat, the middle part of the bent nanoribbon undergoes a large deformation. This makes the SOC energies of the W atoms on the left side and the middle two W atoms of the bent nanoribbon largely increased, while those of the W atoms on the right side decrease slightly, compared to the flat case (SI Figure S4). The asymmetrical change of the SOC energies crossing the ribbon width at the large bending may be related to asymmetrical structure of the nanoribbon, i.e., it is not symmetrical about the plane bisecting the width of the nanoribbon. Those effects induce a strong band splitting effect among the edge bands around the Fermi level and near the Γ point. The large bending also results in a strong band mixing between edge and bulk bands. Similar edge-bulk band mixing is usually observed in other TMD nanoribbons when the edge bands merge into VBC under large bending curvatures [22, 23]. The bands around the Fermi level also show a complex and mixed spin-polarization, as shown in Figures 3(u)-(x). All these features demonstrate that the Z14 at R = 9 Å is topologically trivial.

The changing trend of the Z14 nanoribbon with bending curvatures shows a means for controlling the topological nature of the nanoribbon. The reproducibility and reversibility of bending can achieve the on-off switching of the topological states in the nanoribbon-based quantum devices.

**Optical absorption and exciton states of semiconducting 1T′ WSe$_2$ nanoribbons.** We calculated the band structures of the flat NZ$n$ 1T′ WSe$_2$ nanoribbons with $n$=11, 12, 13, 14, 15, 16, 17, and 18. The relaxed structures and calculated band structures of those NZ$n$ 1T′ WSe$_2$ nanoribbons are shown in SI Figures S1 and S3. Those NZ$n$ 1T′ WSe$_2$ nanoribbons are semiconducting and the band gap shows a non-monotonic changing trend with the width of nanoribbon, with a small gap about 29 meV for NZ14. This width dependent gap feature is due to the quantum confinement effect along the width direction of the nanoribbon. We also study the optical properties and exciton states of the NZ11 1T′ WSe$_2$ nanoribbon under different bending conditions. The relaxed structures of the NZ11 nanoribbon under different bending conditions are shown in SI Figure S2. The flat nanoribbon has almost the same structure as that of the monolayer, only having slight deformations for atoms near the two edges. At R = 7.9 Å, the relative positions of atoms are also almost the same as that in flat case, while the lower Se atom layer experiences substantial compression in the ribbon width direction. At R = 4.5 Å, the ribbon is significantly bent, and the lower Se atom layer further compressed, and the lower two end Se atoms are slightly pushed out and are positioned lower than other atoms. The DFT band structures, $G_0W_0$ band structures, and spin-polarization resolved $G_0W_0$ band structures for the NZ11 nanoribbon under the three bending conditions are shown in Figure 4. For the flat

NZ11, both DFT and $G_0W_0$ show a direct band gap at a k point, denoted as H in Figure 4(a) and (b), where the length of $\Gamma - H$ is about one quarter of that of $\Gamma - Z$, with a slightly larger length of $\Gamma - H$ in the $G_0W_0$ band structure than that in the DFT one. The DFT band gap is 0.21 eV, while the $G_0W_0$ one is 1.16 eV, showing a large self-energy correction to quasiparticle band gap from the many-body perturbation method. The bands around the Fermi level have contributions both from the W atom d-orbital and the Se atom p-orbital, while the edge W atoms' contribution is very small, indicated by the DOS plot in the SI Figure S5. At R = 7.9 Å, both DFT and $G_0W_0$ show an indirect band gap between k points $\Gamma$ and Z for this bent nanoribbon. The band gap is increased from 0.14 eV in DFT to 1.10 eV in $G_0W_0$. For R = 4.5 Å, the band gap is still indirect and is increased from 0.35 eV in DFT to 1.33 eV in $G_0W_0$. Similarly, in the two bent cases, the bands around the Fermi level are also mainly contributed by the W atom d-orbital and the Se atom p-orbital, and the edge W atoms' contribution is still low(SI Figure S5 for the DOS plots).

The calculated optical absorption and exciton spectra are shown in Figure 5. The absorption spectrum of the flat NZ11 nanoribbon shows a strong peak at about 0.3 eV. This is mainly due to the direct band gap at the k-point H. There are also several low magnitude peaks within the range of 05-0.9 eV as can be seen in the enlarged plot in Figure 5(b). The first three excitons with the lowest energy are dark excitons. The first one is at 0.20 eV and is mainly due to transitions V1/V2 to C1/C2 around the k-point H. Since V1/V2 and C1/C2 are of mixed spin-polarization, this dark exciton consists of electron-hole pairs of mixed up-spin and down-spin, hence is of a mixed spin configuration. The bands around the Fermi level have a significant bulk atom contribution and the edge atoms' contribution is negligible. This results in the exciton wavefunction of this dark exciton mainly located in the middle region of the nanoribbon, not on the edges. The wavefunction plots of this dark exciton in the real and momentum spaces and the band-transition composition are shown in Figure 6. The two dark excitons at 0.20 eV and 0.21 eV are mainly resulting from transitions from V2 to C1 and V1 to C2 around the k-point H, as shown in SI Figures S6 and S7. V2 and C1 have opposite spin-polarization around H, so as for V1 and C2. This results in mainly like-spin transitions from hole bands to electron bands for these two dark excitons. Note that the hole has an opposite spin to that of the electron, which was removed from the hole spot to create the hole. Also, the two dark excitons, like the first one, show middle-region distributions of the exciton wavefunctions in the nanoribbon. Their wavefunctions all extend over several unit cells along the ribbon length direction, showing a non-Frenkel feature. The bright exciton with the highest oscillator strength in peak A is at 0.29 eV and is mainly from transitions from V1 to C1 and V2 to C2 around H, with a majority contribution from V2 to C2, (SI Figure S8). Since V2 and C2 have the same spin-polarization around $\Gamma$ and H, as for V1 and C1, this results in a mainly unlike-spin configuration of electron and hole pairs for this bright exciton. This exciton is also spatially located in the middle region of the ribbon. The bright exciton with highest strength in peak B is at 0.52 eV and is due to transitions of V3/V4 to C1/C2 around the $\Gamma$ point, (SI Figure S9). This bright exciton has a mixed spin configuration in its electron-hole pairs and also shows a spatial location of wavefunction in the middle region of the ribbon. The bright exciton at 0.65 eV in peak C is from transitions from V1-V4 to C1-C4 mainly around H and its spatial wavefunction is located in the middle region of the ribbon and show slightly nodal feature, (SI Figure S10). The bright exciton at 0.72 eV in peak D is mainly due to transitions from V5 to C2 and V6 to C1 around $\Gamma$ and it is spatially located in the middle region of the ribbon, see SI Figure S11. The bright exciton at 0.85 eV in peak E is mainly due to transitions of V1 to C4 and V2 to C3, and its wavefunction shows nodal features both in the real space and k space, as shown in SI Figure S12.

For bending at R = 7.9 Å, the first three excitons become bright, and they form peak A and are mainly from transitions of V1/V2 to C1/C2 around $\Gamma$, (SI Figures S13-S15). The conduction bands C1 and C2 are almost flat in the Brillouin zone (BZ), while V1 and V2 are also flat near $\Gamma$ and show a large dispersion (~0.5 eV) throughout BZ. Those bright excitons have a mixed spin configuration and spatially locate in the middle region of the ribbon. The typical bright exciton at 0.24 eV in peak B is due to transitions from V1/V2 to C1-C4 around k points slightly away from the Γ point, (SI Figure S16). This exciton is also of mixed spin configuration and spatially located in the middle of the ribbon with a slight nodal feature. The bright exciton with the highest strength in peak C is at 0.32 eV and also results from transitions from V1/V2 to C1-C4 around k points close to and slightly away from Γ, (SI Figure S17). Its wavefunction is located in the middle region of ribbon and shows a slightly nodal feature. The bright exciton at 0.46 eV in peak D is due to transitions from V1-V4 to C1-C4 around Γ and the k point region away from Γ, see SI Figure S18. It shows nodal features both in the real space and k space. The bright exciton at 0.79 eV in peak E is from broad transitions from V1-V6 to C1-C6 (see SI Figure S19) and it shows nodal features both in the real and k spaces, typical for high energy exciton states.

For bending at R = 4.5 Å, the optical spectrum shows a strong absorption peak A, like the flat case, followed by several weak peaks. The lowest exciton state is a bright state, which is at 0.35 eV and mainly from the transition of V1 to C1 around Γ, (SI Figure S20). This exciton is mainly of unlike-spin configuration of electron-hole pairs, since the V1 and C1 bands are all spin-down polarized at the Γ point. Its wavefunction is spatially located in the middle region of the ribbon. The bright exciton with the highest strength at 0.41 eV in peak A is due to transitions from V1/V2 to C1/C2 around k points Γ and Z, and it is spatially located in the middle region of the bent ribbon, (SI Figure S21). Peaks B, C and D are generally broad and consists of many bright exciton states. Those exciton states are usually formed from a a broad transition and usually show nodal features both in the real and k spaces. For examples, the exciton at 0.54 eV in peak B is due to transitions from V1-V4 to C1-C8. The exciton at 0.70 eV in peak C is also due to transitions from V1-V4 to C1-C8. The exciton at 0.89 eV in peak D is also arises from V1-V4 to C1-C8. These excitons all show nodal features both in the real and k spaces, as shown in SI Figures S22-S24.

A dramatically changed optical absorption spectrum is seen when applying bending R = 4.5 Å to the flat NZ11 nanoribbon, where the absorption edge changed from about 0.20 eV in the flat case to 0.30 eV in the bending case. The strong absorption peak position is also changed by about 0.10 eV. While applying a moderate bending of R = 7.9 Å, the absorption extends to low energy regions. Bending turns the lowest energy exciton states from dark in the flat nanoribbon case into bright in the bent cases. These interesting properties in the semiconducting 1T′ WSe2 nanoribbons will find applications in optoelectronic devices and exciton-based quantum information processes.

**Conclusion**

In summary, we assess the correlation between topology and optical properties in the 1T′ WSe2 nanoribbons under bending from first-principles density functional theory (DFT) and many-body perturbation GW and Bethe-Salpeter equation (GW+BSE) methods. The Z$n$ nanoribbons, with their ribbon length parallel to the zigzag W atom chains and $n$ representing the number of W atoms in the supercell of the nanoribbon, show edge W atom derived edge bands crossing through the Fermi level, while the NZ$n$ nanoribbons with their ribbon length perpendicular to the zigzag chains are semiconducting. The Z$n$ 1T′ WSe2 nanoribbons can show topological edge states with widths of ~4-6 nm. Those edge bands cross the Fermi level an odd number of times and connect the valence and conduction band continua. Each edge band shows one kind of spin-polarization, consistent with spin momentum locked helical edge states. Small and medium bending of R ≥ 12 Å in the nanoribbon do not affect the topological features of the edge bands, showing the robustness of the topological features over these bending deflections. Large bending R = 9 Å generally induces large band splitting around the Fermi level and shift more bulk band contribution to the Fermi level, leading to a switch-off of the topological features in the edge bands by the large bending. The optical absorption and exciton states show a large tunability with bending in the semiconducting 1T′ WSe2 nanoribbons and reflect the spin polarization of the topology. The onset of an absorption edge can shift to the high frequency direction by 0.10 eV when a large bending is applied to the nanoribbon. Bending turns the lowest-energy dark exciton into an optically bright one. These interesting properties in the 1T′ WSe2 nanoribbons indicate potential applications in controllable quantum electronic nanodevices and exciton-based quantum information processes.

**Methods**

Density functional theory (DFT) calculations were conducted in the Vienna Ab initio Software Package (VASP) [27] with projector augmented wave (PAW) pseudopotentials [28, 29]. The Perdew-Burke-Ernzerhof (PBE) [30] functional with SOC was used to calculate the band structures of nanoribbons. The vacuum layer of more than 12 Å is added along the direction of nanoribbon width and inserted along the direction perpendicular to the 2D surface of the nanoribbon, to avoid the interactions between the nanoribbon and its periodic images. The energy cutoff is 500 eV. The k-point mesh of $1 \times 1 \times 100$ was used for band structure calculations. The nanoribbons were built from the pre-relaxed 1T′ WSe2 monolayer, which was relaxed with PBE+SOC and has the in-plane lattice constants $a = 3.302$ Å and $b = 5.941$ Å. All nanoribbons were fully structurally relaxed with all forces less than 0.01 eV/Å. During the relaxation, the x and y coordinates of the two outermost metal atoms on the two edge sides were fixed, while their coordinates along the ribbon axis direction, which is the z direction, and all the coordinates of other atoms were allowed to relax. The supercell vector c along the z axis was also allowed to relax. The $G_0W_0$ [24, 25] and $G_0W_0$+BSE [26] calculations were conducted in BerkeleyGW [24] by pairing with Quantum ESPRESSO [31]. The wavefunction energy cutoff is 70 Ry (~950 eV). The energy cutoff for the epsilon matrix is 18 Ry (~240 eV). The k-point mesh of $1 \times 1 \times 36$ and both valence and conduction bands of 6 was set for optical absorption calculations. The band number for summation is 1100. The correction of the exact static remainder and the wire Coulomb truncation for 1D systems were also used.

**Acknowledgement**

This material is based upon work supported by the U.S. Department of Energy, Office of Science, Office of Basic Energy Sciences, under Award Number DE-SC0021263. The work of H.T. was in part funded through the Catalytic Collaborative Funding Initiative sponsored by the Office of Vice President for Research at Temple University. This research used resources of the National Energy Research Scientific Computing Center, a DOE Office of Science User Facility supported by the Office of Science of the U.S. Department of Energy under Contract No. DE-AC02-05CH11231.

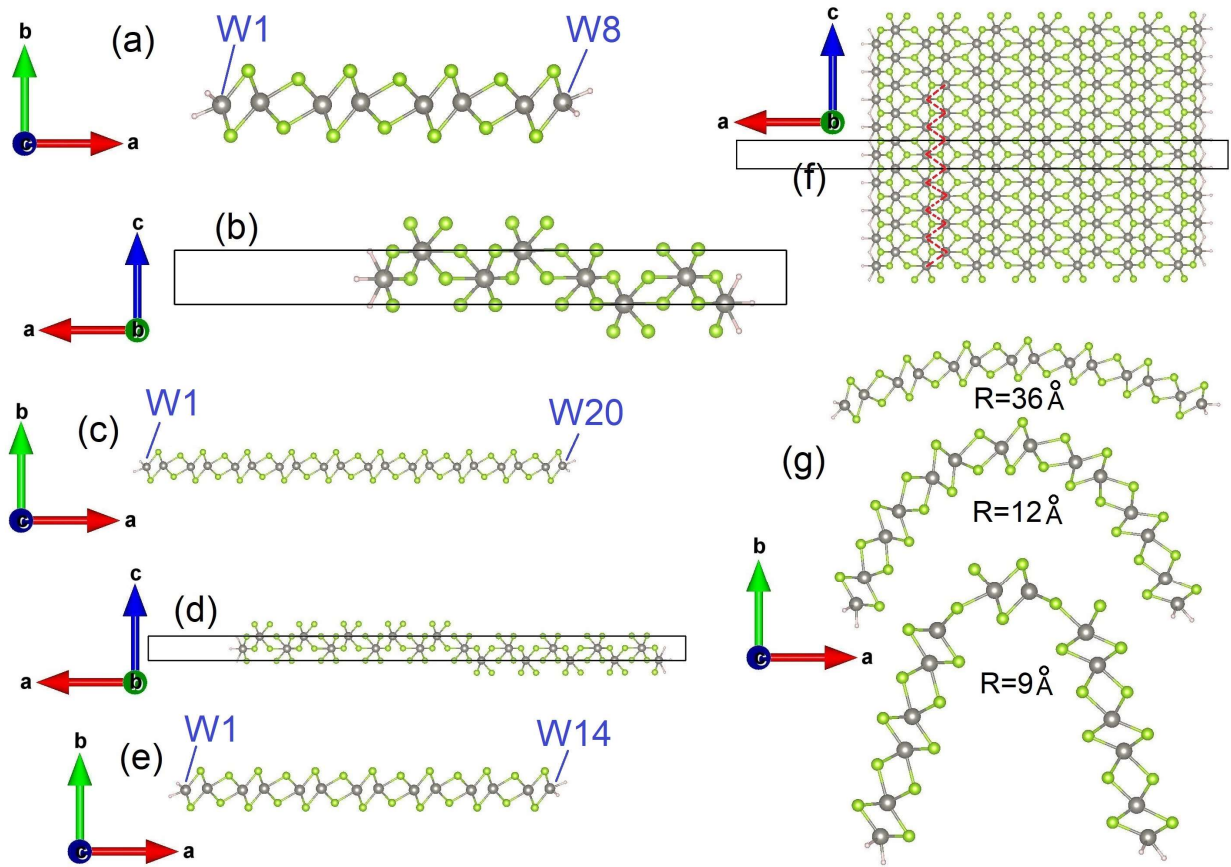

Figure 1. The structures of the 1T′ WSe$_2$ nanoribbons. (a) and (b) are the side and top views of the Z8 nanoribbon, in which the ribbon length direction is aligned along the vector c of the supercell and is parallel to the W atom zigzag chains as shown in (f). (c) and (d) are the two views of the Z20 nanoribbon. (e) is the side view of the Z14 nanoribbon and (f) is for its extended top view. The supercell vectors a, b and c are aligned with the coordinate axes x, y, and z, respectively. The zigzag W atom chain is indicated by the red dashed line in (f). (g) shows the relaxed structures of the Z14 nanoribbon under three bending curvature radii R = 36 Å, R = 12 Å, and R = 9 Å. The green, grey and small white balls represent Se, W and H atoms, respectively. The rectangles in (b), (d) and (f) are the supercells. The two edge W atoms are indicated by the blue letter and numbers in (a), (c) and (e). The graphs in these panels are not plotted in the same scale.

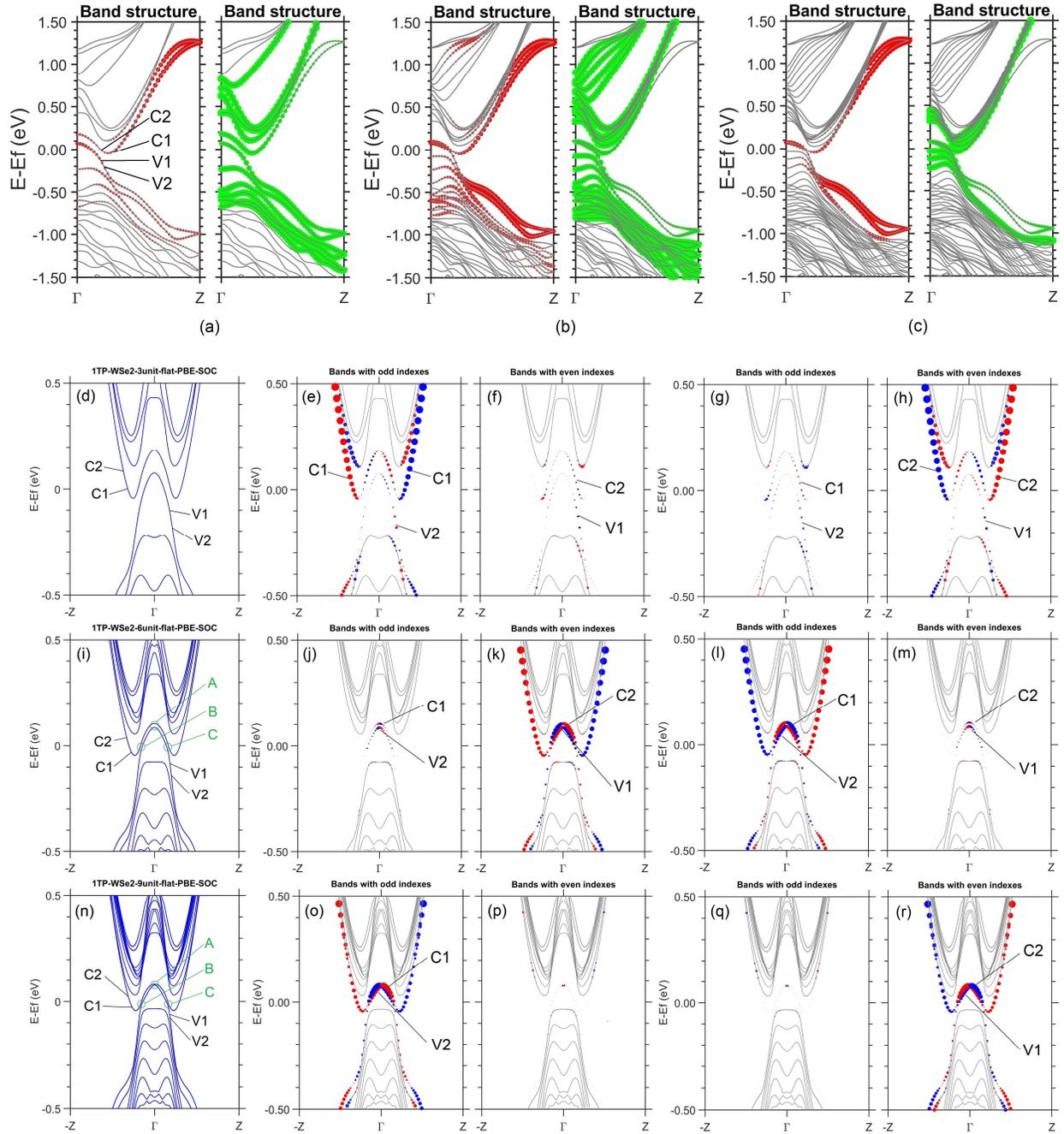

Figure 2. The band structures with edge and bulk atom contributions and spin-polarization resolutions of the flat Z$n$ nanoribbons. In (a), the left panel is the band structure of the flat Z8 with the two edge W atom contributions highlighted by red dots, while the right panel is for other bulk atom contributions highlighted by green dots. (b) and (c) are similarly arranged and for the flat Z14 nanoribbon and the flat Z20 nanoribbon, respectively. The second row of panels (d)-(h) is for spin-polarization resolved band structure around the Fermi level for the flat Z8 nanoribbon. (d) is the ordinary band structure. (e) and (f) are for the spin-polarization resolved band structure with contribution from edge atom W1 highlighted by red and blue dots, where the spin-polarization (the red and blue dots) contributed from atom W1 to the odd-number indexed bands is shown in (e), while the spin-polarization (the red and blue dots) contributed from atom W1 to the even-number indexed bands is shown in (f). Red (blue) dots represent up-spin (dpwn-spin). The mainly

bulk-atom contributed bands are shown as the grey lines in (e) and (f). Panels (g) and (h) are for the contribution from the other edge atom W8 and arranged in the same way as (e) and (f). The third row of panels (i)-(m) is for the flat Z14 nanoribbon and arranged in the same way as the second row, except that (l) and (m) are for the other edge atom W14 in the Z14 nanoribbon. The fourth row of panels (n)-(r) is for the flat Z20 nanoribbon and arranged in the same way as the second row, except that (q) and (r) are for the other edge atom W20 in the Z20 nanoribbon. See Figure 1 for the locations of the edge W atoms.

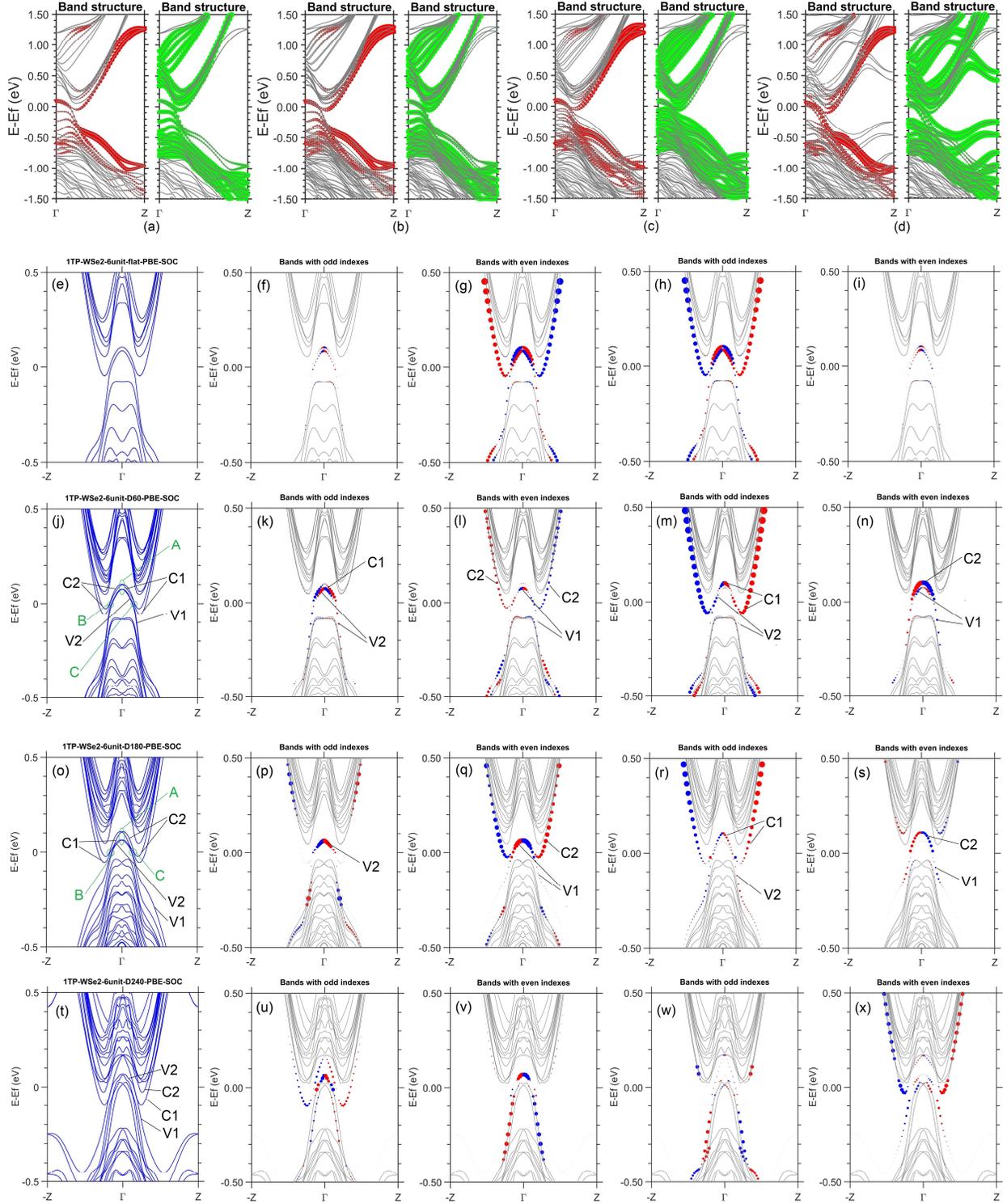

Figure 3. The band structures with edge and bulk atom contributions and spin-polarization resolutions of the Z14 nanoribbon under different bending conditions. In (a), the left panel is the band structure of the flat Z14 with the two edge W atom contribution highlighted by red dots, while the right panel is for other bulk atom contribution highlighted by green dots. (b), (c) and (d) are similarly arranged and for the Z14 nanoribbon under bending curvature radii R = 36 Å, R = 12 Å, and R = 9 Å, respectively. The second row

of panels (e)-(i) is for spin-polarization resolved band structure around the Fermi level for the flat nanoribbon. (e) is the ordinary band structure. (f) and (g) are for the spin-polarization resolved band structure with contribution from edge atom W1 highlighted by red and blue dots, where the spin-polarization (the red and blue dots) contributed from atom W1 to the odd-number indexed bands is shown in (f), while the spin-polarization (the red and blue dots) contributed from atom W1 to the even-number indexed bands is shown in (g). The mainly bulk atom contributed bands are shown as the grey lines in (f) and (g). Panels (h) and (i) are for the contribution from the other edge atom W14 and arranged in the same way as (f) and (g). The third row of panels (j)-(n) is for the Z14 nanoribbon at the bending curvature radius $R = 36$ Å and arranged in the same way as the second row. The fourth row of panels (o)-(s) is for the Z14 nanoribbon at the bending curvature radius $R = 12$ Å and arranged in the same way as the second row. The fifth row of panels (t)-(x) is for the Z14 nanoribbon at the bending curvature radius $R = 9$Å and arranged in the same way as the second row. See Figure 1 for the locations of the edge W atoms.

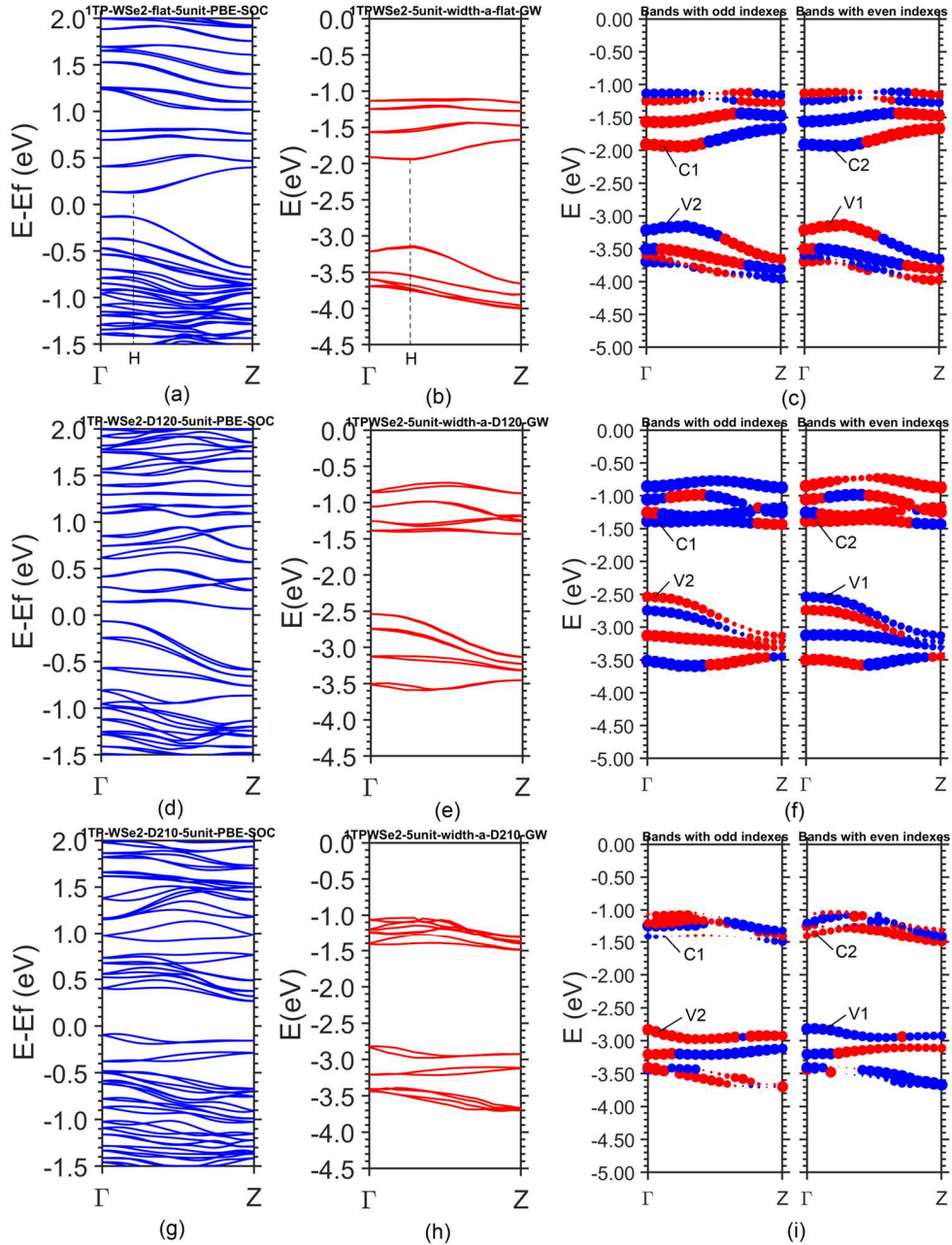

Figure 4. The DFT, $G_0W_0$, and spin-polarization resolved $G_0W_0$ band structures for the 1T′ NZ11 WSe$_2$ nanoribbon under three different bending conditions. Panels (a)-(c) are for the flat NZ11. (a) is the DFT band structure. (b) is the $G_0W_0$ band structure with PBE as the mean field reference. (c) is the spin-polarization resolved $G_0W_0$ band structure of panel (b). In (c) the bands with the odd number and even number indexes are plotted on the left and right subplots respectively for clarity. The second row of plots is for bending curvature radius R = 7.9 Å and the third row is for R = 4.5 Å. The second and third rows are plotted and arranged in the same way as the first row.

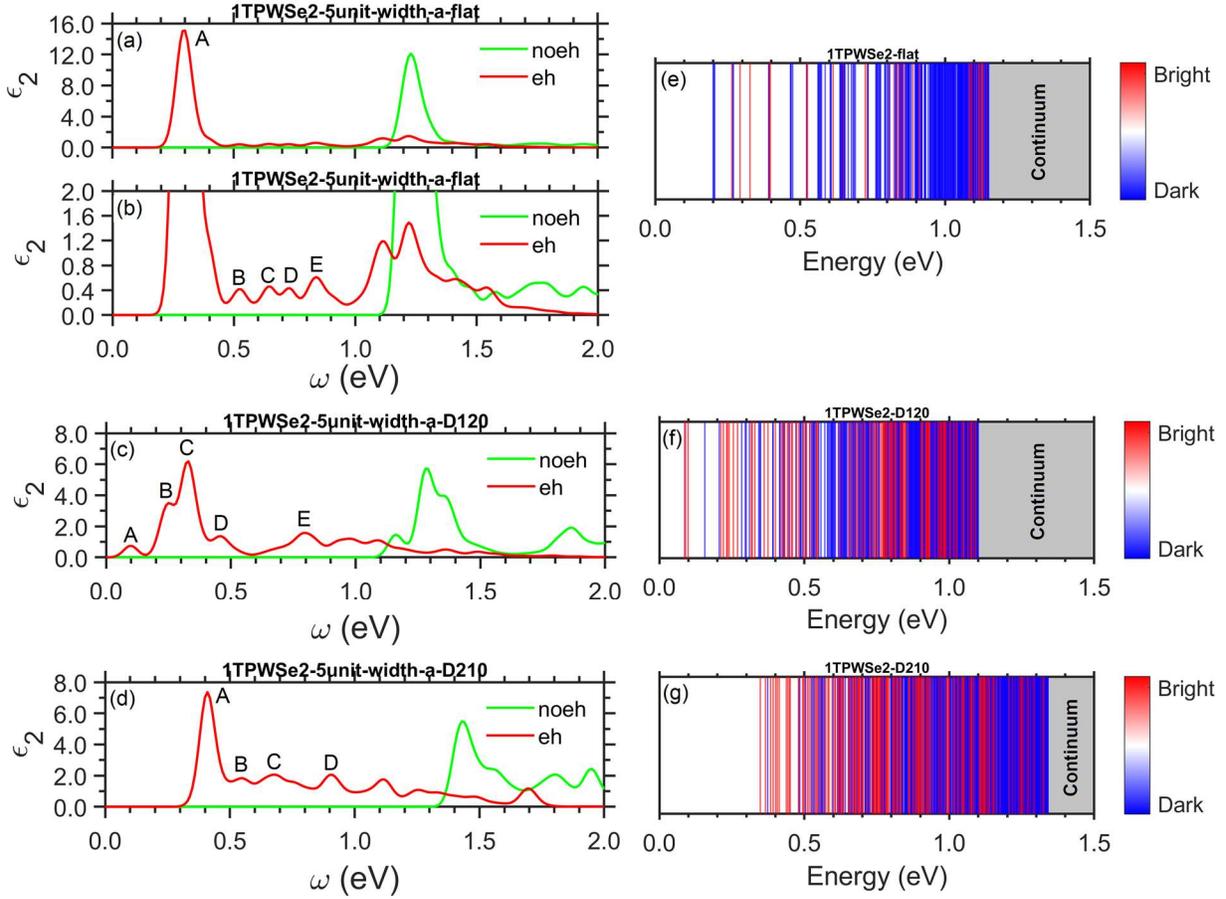

Figure 5. The optical absorption and exciton spectra for the 1T′ NZ11 WSe$_2$ nanoribbon under three different bending conditions. Panel (a) shows the optical absorption spectrum for the flat nanoribbon and is plotted as the imaginary part of the dielectric function as a function of photon energy with the red curve representing the G$_0$W$_0$+BSE result with electron-hole (eh) interactions and the green one for that without eh (noeh) interactions, both with constant broadening of 28 meV. Panel (b) is the same as (a), but with a reduced vertical scale for better details. Panel (e) is the exciton spectrum of the flat nanoribbon. Panels (c) and (f) are similarly plotted and arranged for the bent nanoribbon with R = 7.9 Å. Panels (d) and (g) for the bent nanoribbon with R = 4.5 Å. Bright (dark) exciton states are represented by red (blue) lines in panels (e)-(g).

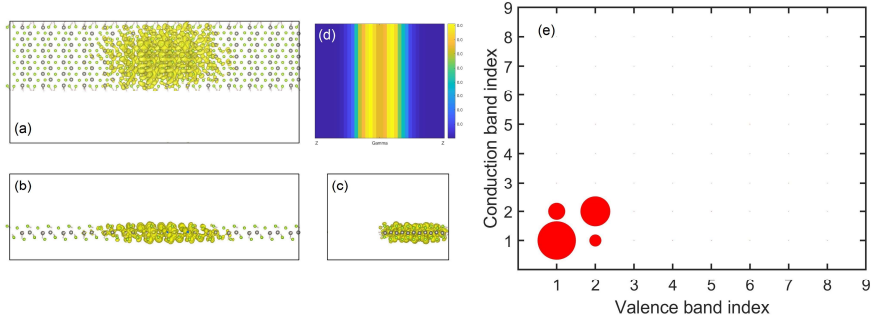

Figure 6. The first dark exciton at 0.20 eV of the NZ11 WSe$_2$ nanoribbon. Panels (a), (b) and (c) represent the three views of the isosurface contour of the squared modulus of the exciton wavefunction in real space, where the hole (black spot in (a)) is located at the center of the ribbon and near a W atom. The profile of the modulus squared exciton wavefunction in k space is shown in panel (d). Panel (e) shows the contributing hole (valence) and electron (conduction) bands for this exciton. The valence (conduction) band index is counted downwards (upwards) from the Fermi level. The spot size in (e) is proportional to $\sum_k |A_{v,c}(k)|^2$ and represents the contributing weight from the v-c pair.